# Towards establishing best practice in the analysis of hydrogen and deuterium by atom probe tomography

Baptiste Gault[1,2]*, Aparna Saksena[1]*, Xavier Sauvage[3], Paul Bagot[4], Leonardo S. Aota[1], Jonas Arlt[5], Lisa T. Belkacemi[6,7], Torben Boll[8], Yi-Sheng Chen[9,10], Luke Daly[4,9,11], Milos B. Djukic[12], James O. Douglas[2], Maria J. Duarte[1], Peter J. Felfer[13] Richard G. Forbes[14], Jing Fu[15], Hazel M. Gardner[16], Ryota Gemma[17,18], Stephan S.A. Gerstl[19], Yilun Gong[1,2], Guillaume Hachet[1], Severin Jakob[20], Benjamin M. Jenkins[3], Megan E. Jones[21], Heena Khanchandani[22], Paraskevas Kontis[20], Mathias Krämer[1], Markus Kühbach[23], Ross K.W. Marceau[24], David Mayweg[20], Katie L. Moore[25], Varatharaja Nallathambi[1,26], Benedict C. Ott[13], Jonathan D Poplawsky[27], Ty Prosa[28], Astrid Pundt[29], Mainak Saha[30], Tim M. Schwarz[1], Yuanyuan Shang[31], Xiao Shen[32], Maria Vrellou[33], Yuan Yu[34], Yujun Zhao[35], Huan Zhao[36], Bowen Zou[37]

[1] Max Planck Institute for Sustainable Materials, Max-Planck-Straße 1, Düsseldorf, Germany

[2] Department of Materials, Imperial College London, Royal School of Mines, Prince Consort Rd, South Kensington, London SW7, UK

[3] Univ Rouen Normandie, INSA Rouen Normandie, CNRS, Groupe de Physique des Matériaux, UMR6634, 76000 Rouen, France

[4] Department of Materials, University of Oxford, Parks Road, Oxford OX1 3PH, United Kingdom

[5] University of Göttingen, Institute for Materials Physics, Göttingen D-37077, Germany

[6] Leibniz-Institute for Materials Engineering-IWT, Badgasteiner Straße 3, 28359 Bremen, Germany

[7] MAPEX Center for Materials and Processes, Universität Bremen, Bibliothekstraße 1, 28359 Bremen, Germany"

[8] Institute for Applied Materials (IAM-WK) and Karlsruhe Nano Micro Facility (KNMFi), Karlsruhe Institute of Technology (KIT), Hermann-von-Helmholtz-Platz 1, D-76344, Eggenstein-Leopoldshafen, Germany

[9] Australian Centre for Microscopy and Microanalysis, The University of Sydney, Australia

[10] School of Materials Science and Engineering, Nayang Technological University, Singapore

[11] School of Geographical and Earth Sciences, University of Glasgow, Glasgow, UK

[12] University of Belgrade, Faculty of Mechanical Engineering, Kraljice Marije 16, Belgrade 11120, Serbia

[13] Department of Materials Science & Engineering, Institute I: General Materials Properties, Friedrich-Alexander-Universität Erlangen-Nürnberg, Erlangen, Germany

[14] Quantum Foundations and Technologies Group, School of Mathematics and Physics, University of Surrey

[15] Department of Mechanical and Aerospace Engineering, Monash University, Melbourne, Australia

[16] Materials Science and Engineering, UK Atomic Energy Authority, Culham Science Centre, Abingdon, Oxfordshire, UK.

[17] Dept. of Applied Chemistry, Tokai University, 259-1292, Kanagawa, Japan

[18] Micro/Nano Technology Center, Tokai University, 259-1292, Kanagawa, Japan

[19] Scientific Center for Optical and Electron Microscopy, ETH Zurich, 8093 Zurich, Switzerland

[20] Department of Physics, Chalmers University of Technology, SE-412 96 Göteborg, Sweden

[21] National Nuclear Laboratory, Windscale Laboratory, Sellafield, Seascale, Cumbria, CA20 1PG, United Kingdom

[22] Department of Materials Science and Engineering, Norwegian University of Science and Technology, Trondheim 7491, Norway

[23] Consortium FAIRmat, Humboldt-Universität zu Berlin, Zum Großen Windkanal 2, D-12489 Berlin, Germany

[24] Deakin University, Institute for Frontier Materials, Geelong Waurn Ponds Campus, Waurn Ponds, VIC 3216, Australia

[25] University of Manchester, Oxford Road, Manchester, M21 0XT, UK

[26] Technical Chemistry I and Center for Nanointegration Duisburg-Essen (CENIDE), University of Duisburg-Essen, 45141 Essen, Germany

[27] Center for Nanophase Materials Sciences, Oak Ridge National Laboratory, Oak Ridge, TN 37830, USA

[28] CAMECA Instruments, Inc. 5470 Nobel Drive in Madison, WI 53711 USA

[29] Karlsruhe Institute of Technology KIT, IAM-WK, Kaiserstraße 12, 36131 Karlsruhe

[30] Research Centre for Magnetic and Spintronic Materials, National Institute for Materials Science, Sengen, Tsukuba, Ibaraki-305-0047, Japan

[31] Department of Materials Design, Institute of Hydrogen Technology, Helmholtz-Zentrum Hereon GmbH, 21502, Geesthacht, Germany

[32] Institute of Materials Engineering, University of Kassel, Moenchebergstr.3 34125 Kassel

[33] Institute for Applied Materials, Karlsruhe Institute of Technology, Kaiserstrasse 12, 76131 Karlsruhe, Germany

[34] Institute of Physics (IA), RWTH Aachen University, 52056 Aachen, Germany. E-mail: yu@physik.rwth-aachen.de

[35] Institute for Materials, Ruhr-Universität Bochum, Universitätsstraße 150, 44801 Bochum, Germany

[36] State Key Laboratory for Mechanical Behavior of Materials, Xi'an Jiaotong University, Xi'an 710049, China

[37] Institute of Materials Engineering, University of Kassel, Möncheberstraße 3, 34125 Kassel, Germany

* Corresponding author: b.gault@mpie.de | a.saksena@mpie.de

## Abstract

As hydrogen is touted as a key player in the decarbonization of modern society, it is critical to enable quantitative H analysis at high spatial resolution – if possible at the atomic scale. Indeed, H has a known deleterious impact on the mechanical properties (strength, ductility, toughness) of most materials that can hinder their use as part of the infrastructure of a hydrogen-based economy. Enabling H mapping, including local hydrogen concentration analyses at specific microstructural features, is essential for understanding the multiple ways that H affect the properties of materials, including for instance embrittlement mechanisms and their synergies, but also spatial mapping and quantification of hydrogen isotopes is essential to accurately predict tritium inventory of future fusion power plants, ensuring their safe and efficient operation for example. Atom probe tomography (APT) has the intrinsic capabilities for detecting hydrogen (H), and deuterium (D), and in principle the capacity for performing quantitative mapping of H within a material's microstructure. Yet the accuracy and precision of H analysis by APT remain affected by the influence of residual hydrogen from the ultra-high vacuum chamber that can obscure the signal of H from within the material, along with a complex field evaporation behavior. The present article reports the essence of discussions at a focused workshop held at the Max-Planck Institute for Sustainable Materials in April 2024. The workshop was organized to pave the way to establishing best practices in reporting APT data for the analysis of H. We first summarize the key aspects of the intricacies of H analysis by APT and propose a path for better reporting of the relevant data to support interpretation of APT-based H analysis in materials.

# 1 Introduction

Hydrogen is the smallest and lightest atom, and is most abundant, making it ubiquitous. Within the microstructure of materials, hydrogen is known to cause a drop in toughness, ductility, and resistance to crack propagation through an array of possible mechanisms referred to as hydrogen embrittlement (Sofronis & Robertson, 2006; Pundt & Kirchheim, 2006; Hirth, 1980; Robertson et al., 2015; Lynch, 2019). Despite decades of research, there are still open questions as to the active mechanism(s), and how to identify them, which is a prerequisite to define strategies to circumvent (or delay) hydrogen embrittlement and enhance the durability and sustainability of engineering parts (Bhadeshia, 2016; Djukic et al., 2019). There are numerous other aspects of hydrogen trapping inside materials, for instance trapped tritium can pose a radiological hazard, especially for maintenance and decommissioning, and that trapped tritium can cause inventory issues by reducing the amount of tritium available for use as fuel.

If scanning tunnelling microscopes have been used to manipulate and address individual H atoms on semiconducting surfaces (Simmons et al., 2003), quantitative H analysis at high spatial resolution, at least sufficient to determine directly the distribution of H across the microstructure of an engineering alloy, remains extremely challenging. Insights are however necessary to complement and inform bulk measurements, either from X-ray or neutron scattering or diffraction (Maxelon et al., 2001), Kelvin probe experiments in a permeation configuration (Evers et al., 2013) or thermal-desorption spectroscopy (TDS) (Merzlikin et al., 2015; Choo & Lee, 1982) for instance. Forays in this direction have been made in secondary-ion mass spectrometry (Aboura & Moore, 2021; Jones et al., 2021), TDS (Suzuki & Takai, 2012), for hydrides, through transmission-electron microscopy (TEM) (de Graaf et al., 2020; Hamm et al., 2019). TEM has also been extensively used for investigating the effects of hydrogen deformation mechanisms (Robertson, 2001), yet with no direct microanalytical capabilities for locating H in the microstructure .

In principle, atom probe tomography (APT) is the only technique that can combine a capacity for direct detection of H with capabilities for nanoscale, three-dimensional mapping (Cerezo et al., 2007; Kelly & Miller, 2007; Gault et al., 2021). The single-particle detector that equips modern APs (Da Costa et al., 2005; Kelly et al., 2004) operate in a uses microchannel plates (MCP) to convert and amplify the signal from the ion impact. The MCPs are operated in a saturated mode that ensures that their efficiency does not depend on the mass or the energy of the incoming ion. The detector can hence detect $H^+$ ions, and the mass resolution is typically sufficient to distinguish $H^+$ from $^2H^+$ or $D^+$ when isotopic labelling is used. There have been numerous reports of using APT to study the nanoscale distribution of H or D in multilayers (Gemma et al., 2007a, 2011), in steels (Takahashi et al., 2010, 2018; Y.-S. Chen et al., 2017; Chen et al., 2020; Khanchandani et al., 2023; Jakob et al., 2024; McCarroll et al., 2022; Liu et al., 2024), Al-based (Freixes et al., 2022; Zhao et al., 2022, 2024) and Ti-based (Joseph et al., 2022; Chang et al., 2018) alloys for instance, as well as to study hydrides (Breen et al., 2018; Mouton et al., 2021; Jones et al., 2022; Mayweg et al., 2023), with numerous other examples in metallic and non-metallic materials (Tweddle et al., 2019; Shi et al., 2022; Martin et al., 2016; Ott et al., 2024), including geological (Liu et al., 2022; Daly, Lee, Darling, et al., 2021) and extraterrestrial materials (Greer et al., 2020; Daly et al., 2020; Daly, Lee, Hallis, et al., 2021).

However, analysis of H by APT appears subject to an extreme sensitivity to experimental details, including the of H-loading samples with hydrogen or deuterium, their preparation and handling pre- and post-H-loading, the parameters used during acquisition and processing of the data (Gemma et al., 2011, 2009; Breen et al., 2020; Chang et al., 2018; Yanhong Chang et al., 2019; Y.H. Chang et al., 2019; Meier et al., 2023; Takahashi et al., 2010, 2018; Jakob et al., 2024). The following sections summarize selected aspects of the discussions raised during the workshop, from the challenges associated with analyzing H by APT, to possible mitigating solutions, to recommendations for best practices for reporting APT data – with an emphasis on H analysis. We finally provide some perspectives on the future of the field.

## 2 Challenges of H analysis by APT

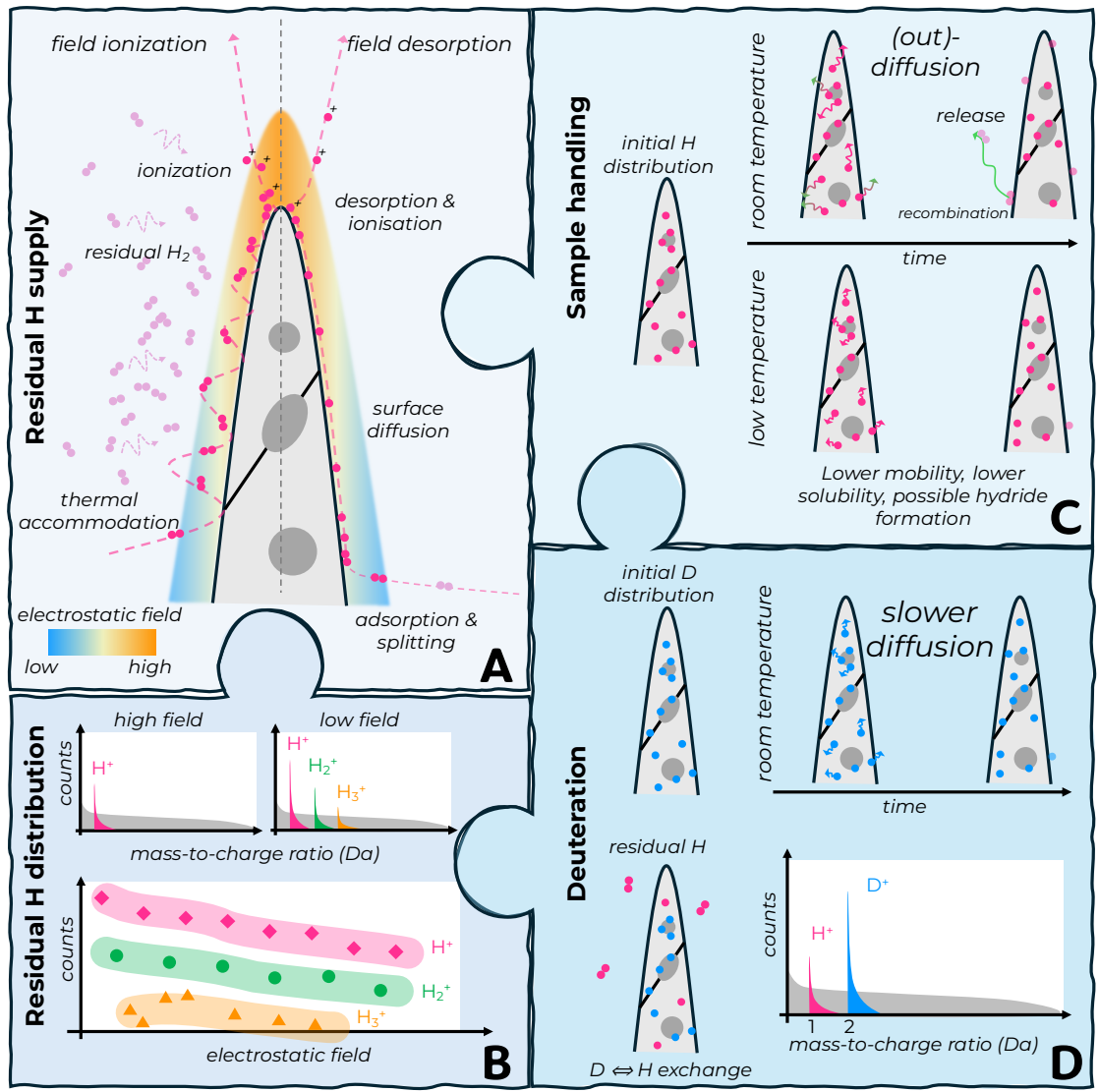

*Figure 1: **A** supply of H to the emitting area at the tip of the specimen from the residual gas; **B** electrostatic field dependence on the ionic distribution of H; **C** high diffusivity of H that can cause losses from the APT specimen during handling at ambient temperature ; **D** advantages of deuteration with regards to slower diffusion, and differentiation between H and D in the mass spectrum.*

Figure 1 recaps some of the main challenges associated with the analysis of H by APT, which were discussed at length over the course of the workshop, and are summarized in the following sections. These include the supply function of residual H (A), the dependence of the distribution of $H^+$, $H_2^+$ and $H_3^+$ on the electrostatic field (B), the need to control the conditions of temperature and pressure to maintain the H within the microstructure (C), and possible advantages pertaining to the use of deuteration to facilitate the analysis of H by APT (D).

## 2.1 Sources and behaviors of residual H

Despite the ultra-high vacuum in the analysis chamber, residual H is most typically detected and can obscure the signal from the ions originating from the specimen. This H comes from the progressive desorption of $H_2$ from the stainless steel of the chamber walls. It is said that the early atom probe chambers made of glass did not face similar issues. Alternative and promising solutions are being actively pursued to limit the amount of residual H, for instance through the use of chambers and parts made of Ti-based alloys instead of the conventional stainless steel (Felfer et al., 2022). This could be coming in addition to the typical getters or cryo-pumping systems more conventionally used.

Once released, gaseous $H_2$ gets attracted to the cold and charged specimen, not unlike the imaging gas dynamics in field-ion microscopy (FIM). The gas supply function in FIM has previously been discussed intensely (Brandon, 1963; Forbes, 1996, Anon, 2008). In short, through a series of hops on the surface, the polarized gas atoms or molecules progressively lose energy and thermalize with the surface, while moving up the electrostatic field gradient towards the apex, Figure 1A. As their velocity becomes lower, and the electrostatic field higher, the probability for ionization increases, with a maximum probability at a so-called critical distance from the surface, as already noted for FIM (Müller & Bahadur, 1956; Forbes, 1996).

An alternative source of H is through the formation of an adsorbed layer. Following thermalization, and assisted by the electrostatic field, some of the gas atoms or molecules chemisorb on the surface itself (Rendulic & Knor, 1967). At the surface, $H_2$ molecules can dissociate into atomic hydrogen. Driven by polarization forces, these atoms or molecules can diffuse along the shank of the specimen towards the apex (Tsong & Kellogg, 1975). The probability of surface diffusion will hence depend strongly on the nature and energy landscape of the surface (Yoo et al., 2022). Through these mechanisms, H can be continuously supplied to the area being analyzed by APT. This raises the question of whether it would be possible to deposit a material specifically along the specimen's shank that could either slow down or even trap, i.e. getter, the H. This could prevent H from reaching the tip of the specimen.

Upon reaching critical electrostatic fields in the near-apex region, the probability of field desorption of the adsorbed H becomes sufficiently high to cause the departure of H either as atomic $H^+$ or as a molecular-ions, i.e. $H_{2-3}^+$, or a metal-hydride-ion, $M_xH_y^{n+}$ for example $TiH_2^+$. There is a reported dependence on the crystallographic facets imaged by FIM (Martinka, 1981) where it can enhance the FIM contrast and resolution, possibly through a modification of the electrostatic field distribution (Müller et al., 1965), and can facilitate field evaporation at relatively lower electrostatic fields (Rendulic & Knor, 1967). These complex mechanisms and fundamental questions remain to be fully clarified. Studies of field evaporation in FIM were mostly performed under electrostatic field conditions, whereas field evaporation in APT is triggered by high-voltage (HV) or laser pulses. Because of the time it takes for the H to migrate along the shank up to the imaged surface at the tip of the needle, the detected amount of hydrogen depends strongly on the time between field evaporation events, which is a combination of laser pulsing frequency and detection rate (Sundell et al., 2013; Kolli, 2017; Meier et al., 2023).

## 2.2 H and field evaporation

H can facilitate field evaporation of surface atoms at a relatively lower electrostatic field (Müller et al., 1965; Wada et al., 1983). These ions can be in the form of hydrogen-bearing molecular ions, i.e. $M_xH_y^{n+}$, as noted in (Krishnaswamy & Müller, 1977), and different metallic elements show a different propensity to form these hydride ions (Stepien & Tsong, 1998), which should not be confused with hydrides as a phase in the thermodynamic sense. They report that Zr and Ti are prone to forming these including in the 2+ charge state, which agrees with more recent reports (Y.H. Chang et al., 2019; Mouton et al., 2018) and was also discussed in the workshop (A. Diagne, CNRS-GPM, Rouen, France). Other hydrogenated ionic species have also been reported (Greer et al., 2020; Heck et al., 2014).

Another complexity is that H can be detected in various forms, i.e. $H^+$, $H_2^+$ and $H_3^+$, along with hydride ions (e.g. $TiH^+$, $TiH_2^+$). The distribution of their relative abundances depends on the strength of the electrostatic field (Tsong et al., 1983), which may need to be assessed for each dataset individually, and even for each of the analyzed microstructural feature of interest within an APT dataset.

There are still many open questions, for instance, how hydrogen would behave with individual substitutional or interstitial elements constituting an alloy of interest? how it is supplied across the field-of-view? Why is detected more at crystallographic poles or some facets? is it subject to surface diffusion also when originating from within the material? (Martinka, 1981; Gemma et al., 2011). If so does it migrate towards protruding particles or precipitates that require a higher electrostatic field to field evaporate (Miller & Hetherington, 1991; Vurpillot et al., 2000), which could artificially increase its concentration near features such as carbides in steels (Breen et al., 2020) for instance or precipitates in Zr-alloys (Jenkins et al., 2023)?

## 2.3 Hydrogen and deuterium

Finally, there are other challenges arising from the high mobility of H in most materials, which can enable diffusion between specimen preparation, H-loading, specimen transfer and even during the APT analysis. This diffusion can to some extent be mitigated by using lower temperatures. The use of a coating on the surface can act as a permeation barrier and help prevent out diffusion of H from the loaded sample (Hollenberg et al., 1995; Kremer et al., 2021). Nevertheless, H can also move or migrate by atomic tunnelling through the lattice, which will limit the efficacy of using low temperatures – see (Gemma et al., 2009) and references therein. It should be noted that, herein, we use diffusion to refer to the motion of hydrogen through the material, which can be effectively related to diffusion or through tunnelling.

Deuteration, i.e. the use of deuterium (D or $^2H$) in lieu of H, offers multiple advantages to facilitate quantitative analysis of H in materials by APT. Deuterium is a heavier isotope of H, with a natural abundance of only 0.0145 at.%. Due to its higher mass, it diffuses more slowly than H and its tunnelling rate is lower (Maxelon et al., 2001). D is hence less prone to moving between specimen preparation and analysis, as well as during the analysis (Gemma et al., 2009). Even at low temperature, i.e. near 25 K, diffusion of H and D can occur. The composition profile for H and D across an $Al_3Zr$ dispersoid reported by Zhao et al (Zhao et al., 2022) shows a tendency for a higher composition of H and D nearer to the freshly exposed surface during the analysis. Although this may be due to a locally higher misfit strain, the profile could be interpreted as a proof of diffusion during the experiment, as raised by Prof. A. Pundt (KIT) during the discussions.

Isotopic marking with D can also alleviate some of the issues arising from the overlap with residual gaseous hydrogen as well. D should in principle be detected at 2 Da and not 1 Da (Gemma et al., 2007b, 2011; Haley et al., 2014; Takahashi et al., 2010). This is provided that the experimental conditions, particularly the intensity of the electrostatic field, are selected to minimize the detection of residual H in the form of $H_2^+$. Otherwise, the overlap between the two signals will not allow for distinguishing between the two ions. Indeed, conversely to NanoSIMS that has sufficient mass resolution, as presented by Dr. K. Moore (University of Manchester) during the workshop (Li et al., 2019), modern APs have insufficient mass resolution. The possibility of designing novel atom probe with sufficient mass resolution was not discussed, but could be an interesting avenue of research in the future.

In some datasets, peak splitting has been observed for the 2 Da peak, which could enable differentiation of $D^+$ and $H_2^{2+}$ ions (Meier et al., 2022). For a D-loaded W sample, analysed using laser assisted evaporation in a straight flight path instrument, a sharp peak is observed at the leading edge of the 2 Da peak, which is not present in the unloaded sample. A broad peak at the trailing edge of the 2 Da peak is present in both the charged and uncharged sample. The number of ions contributing to the broad peak varies as a function of electrostatic field, whereas that of the sharp peak does not. This suggests the broad peak corresponds to the contaminant hydrogen species $H_2^+$, whereas the sharp peak could correspond to $D^+$. Peak splitting is not observed when the same D-loaded sample is analyzed using a reflectron-fitted atom probe, suggesting that the contaminant $H_2^+$ exhibits with an energy deficit causing a difference in the time-of-flight that gets corrected by the reflectron.

Additionally, the field evaporation behavior of H-containing materials may however lead to complexities in the elemental or ionic identification of the peaks in the mass spectrum due to the overlap of hydride- and deuteride ions, e.g. $ZrH_2^+$ and $ZrD^+$ (Mouton et al., 2018; Jones et al., 2022), particularly in cases where a substantial fraction of the signal at 2 Da can be attributed to $H_2^+$.

It also often appears that following D-loading, a peak at 3 Da appears, despite the electrostatic field conditions selected such that H should be detected primarily as $H^+$ at 1 Da. This could be associated to $HD^+$ and incomplete deuteration, i.e. the solution or gas used for H-loading contained a fraction of H, or to exchange between H and D during storage for instance. The exact mechanisms, along with those responsible for other phenomena observed during APT analysis of H, remains largely unexplained and will require further studies.

# 3 Optimizing H analysis by APT

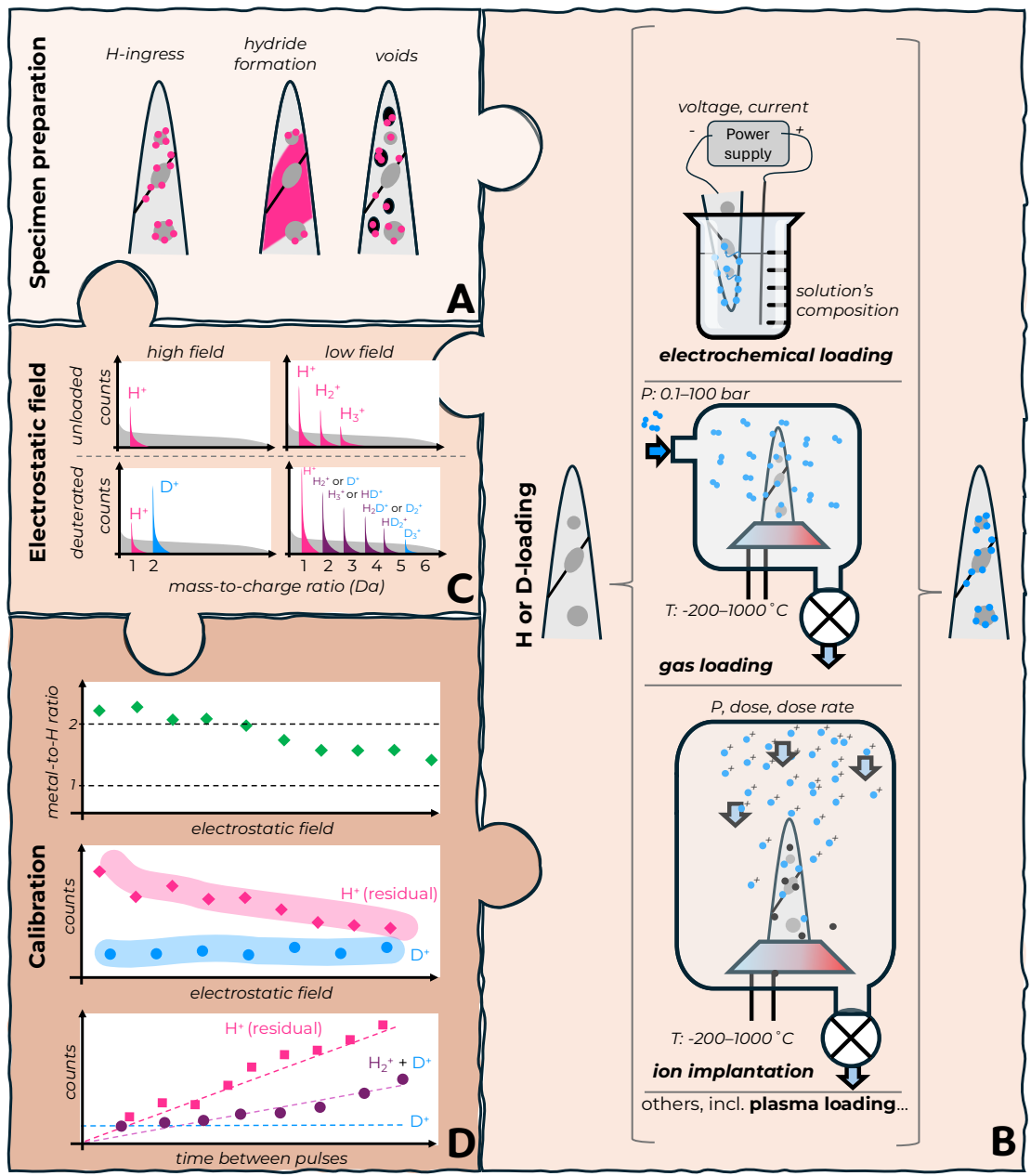

*Figure 2: Summary of the complexity of the analysis of H by APT: **A** ingress of H and structural damage from specimen preparation; **B** numerous methods for H- or D-loading and their most relevant parameters; **C** complexity of mass spectra obtained in different conditions of the electrostatic field; **D** possible calibrations needed to facilitate interpretation of H/D-analysis by APT.*

Figure 2 summarizes (some of) the key aspects discussed herein and during the workshop. These are some of the pieces of a complex puzzle. The following sections discuss specifically specimen preparation (A), the H-/D-loading methods (B), optimization of the electrostatic field conditions during analysis (C), and possible means and interests of performing calibrations (D).

## 3.1 Specimen preparation

### 3.1.1 General considerations

Specimens for APT must be prepared in the shape of sharp needles, with an end radius below 100 nm. They can be prepared by electrochemical polishing is done from a blank, i.e. a small parallelopiped bar or a piece of wire of the material, typically 400 µm across and 1–3 cm long. An alternative route is to use a focused-ion beam (FIB) microscope, often coupled with a scanning electron microscope (SEM) (Prosa & Larson, 2017), or sometimes a combination of both (Miller & Russell, 2006). It should be noted that the specimen preparation may lead to some degree of stress relaxation, which can lead to e.g. rearrangement of dislocations within the sample, and potentially to hydrogen relocation.

Additionally, there are also approaches to fabricate materials for analysis directly onto pre-shaped specimens after they have been cleaned by field evaporation. Many of the early studies of multilayers by APT were performed on thin films or multilayers deposited directly onto needle-shaped substrates (Larson et al., 2009). This was also the case for H-studies by Gemma et al. (Gemma et al., 2007a, 2012, 2011; Kesten et al., 2002). These are however not the most commonly analyzed samples.

### 3.1.2 H-ingress during preparation

An additional complexity for studying H by APT was the recent reports that point to the ingress of hydrogen during specimen preparation (Chang et al., 2018; Yanhong Chang et al., 2019; Breen et al., 2020; Mouton et al., 2021; Yan et al., 2019). In the case of Ti or Zr alloys, this had been reported in the preparation of specimens for electron microscopy (Banerjee & Williams, 1983; Ding & Jones, 2011; Hanlon et al., 2019), and may have been known to the APT community previously – a comment in this direction was made at the APT&M 2018 conference in Gaithersburg, MD by Prof. G. Schmitz (University of Stuttgart) – but only rarely reported (Parvizi et al., 2014).

Following preparation of specimen from a pearlitic steel by electrochemical polishing, Breen and co-workers reported substantial amounts of H, and the amplitude of the H signal dropped substantially following heating at moderate temperature (<150°C) (Breen et al., 2020). They interpreted this as a proof that H had been introduced into the specimen during the sharpening process because of uncontrolled electrochemical conditions, forming atomic hydrogen on the surface and leading to unintended uptake and H-loading. This H can diffuse into the material and saturate the microstructural traps, thereby preventing efficient D-loading into microstructural traps. In preceding discussions at APT&M conferences, the group at the University of Oxford had reported unpublished results that the preparation of APT specimens, by using fully deuterated solutions throughout the electrochemical polishing process, did not result in a similar deuteration of the specimen (Chen, 2017). This point will hence require additional systematic experiments to be completely clarified.

Evidence has pointed to the introduction of significant amounts of H in Ti and Ti-alloys, and in Zr-alloys during the final stages of specimen shaping during FIB-based specimen preparation at room temperature. In this case, it is assumed that the H originates from the corrosion of the freshly created surface by the residual moisture inside the FIB, which, through water splitting leads to atomic H on the specimen surface. It is also possible that other gaseous species, including hydrocarbon or simply $H_2$, present inside the vacuum chamber of the SEM-FIB are dissociated on the surface, thereby forming atomic H. Once created, atomic H can penetrate inside the sample particularly through phases that have a high solubility for H, e.g. β-Ti (Chang et al., 2018; Yan et al., 2019), or through microstructural defects, e.g. grain boundaries (Chang et al., 2018). The H can accumulate to an extent where a hydride phase (i.e. TiH or $ZrH_2$) forms at room temperature. No such reports exist for steel and ferrous-alloys, but corrosion could also lead to H formation and ingress (Holroyd, 1988; Eliezer et al., 1979; Parvizi et al., 2017). Cryogenic specimen preparation by FIB was reported to prevent (or at least drastically limit) the pick-up of H during FIB-preparation of Ti- and Zr- based alloys (Yanhong Chang et al., 2019; Mouton et al., 2021; Mayweg et al., 2023; Jenkins et al., 2023). This may be related to a combination of moisture being trapped by the cold stage and cold finger or the cryo-stage, as well as with a reduction of the kinetics of water splitting or of the diffusivity of hydrogen.

### 3.1.3 Ion and electron beam damage

Beyond the formation of hydrides, FIB-based preparation was shown to induce potential structural damage, in the form of vacancies (Larson et al., 1999; Miller & Russell, 2006). Considering the high trapping energy of hydrogen by vacancies in most metallic systems (Canto et al., 2014), introducing vacancies during preparation could affect the existing hydrogen distribution. Recently, Saksena and co-workers reported void formation from the agglomeration of FIB-induced vacancies that affected the H-H- capabilities of two-phase steels, as a high fraction of the loaded D was trapped in these voids (Saksena et al., 2023). They offered a possible specimen preparation workflow that minimizes these issues by using electron-beam Pt deposition for protection, and then using a Xe-plasma FIB during sharpening.

A final aspect that was discussed following the presentation of Dr. D. Mayweg (Chalmers University) is the possible need to prepare specimens along specific crystallographic orientations, since the distribution of H or D, along with H- or D-containing molecular ions, is dependent on the crystallographic facets at the end of the specimen (Martinka, 1981; Y.H. Chang et al., 2019). This is particularly important in cases where calibrations are performed that similar orientations are compared. This may require imaging the surface before H-loading by electron-backscattered diffraction (EBSD), or the use of transmission Kikuchi diffraction (TKD) (Babinsky et al., 2014) or TEM (Henjered & Norden, 1983) sequentially during the preparation to ensure that a grain with the appropriate orientation is located at the tip of the specimen. This approach may come with the caveat that structural damage can be induced by the recoil of surface species under electron-beam illumination (Gault et al., 2023). Cold traps inside the microscope or intermediate specimen cleaning by low-energy Ar milling (Herbig & Kumar, 2021) could help reduce the contamination. Structural defects generated from this recoil-associated damage can affect the subsequent distribution of H or D after loading.

### 3.1.4 Specimen transport and vacuum-cryo-transfers

Corrosion leading to the creation and ingress of atomic hydrogen at the surface could also occur during exposure to atmosphere during preparation or transfer of specimens into the atom probe chamber. This is conventionally done through ambient atmosphere, but can be mitigated by the use of vacuum shuttles. These also typically offer the possibility to perform the transfer at cryogenic temperature, achieved by using liquid nitrogen. Such a workflow offers the advantage of limiting the thermally-activated diffusion of trapped H and mobile or trapped D throughout the microstructure. This combination of high- or ultra-high-vacuum levels to prevent frost formation, and cryogenic suitcases is increasingly available in APT facilities across the world (Y.-S. Chen et al., 2017; Chen et al., 2019; Stephenson et al., 2018; McCarroll et al., 2020).

An alternative could be, following H-loading, to protect the surface by using a conformal coating with a material with very limited permeability for H or D, to limit the exchange with the atmosphere during transport. For instance, H or D inside the specimen can reach the surface, recombine and desorb. This can deplete the subsurface region, create a gradient that can drive further diffusion of H or D from the inner part of the specimen towards the surface. A coating could slow down this outgassing, and hence the associated diffusional processes. Conformal coatings of APT specimens have been reported (Seol et al., 2016; Adineh et al., 2018; Kim et al., 2022). Recently this was reported directly in situ the FIB (Schwarz et al., 2024), including at cryogenic temperature (Zhang et al., 2021; Woods et al., 2023), however these are typically metallic in nature and may not offer the necessary impermeability for H or D. Forming an oxide coating by using e.g. an oxygen-based plasma or dosing O into the chamber could be a way to form a conformal coating with a lower permeability (Kremer et al., 2021).

Finally, full cryogenic workflows including (site-specific) specimen preparation by FIB lift-out (Schreiber et al., 2018; Douglas et al., 2023), surface protection, transport into the atom probe to analysis can be performed. However, the use of such workflows have not yet been reported to study the distribution of H or D.

## 3.2  Loading with H or D

### 3.2.1  Loading modes

Loading with H and D has been reported through electrochemistry (Haley et al., 2014), gas-phase (Gemma et al., 2012; Khanchandani, El-Zoka, et al., 2022), or ion implantation (Walck & Hren, 1984; Daly, Lee, Hallis, et al., 2021). The use of a H-rich plasma has also been mentioned (Maier et al., 2019). Note that workflows have been introduced for loading with H, D and also tritium (T), and it is important to keep in mind that they will all behave slightly differently through small differences in adsorption energy and mobility through the material for instance.

The optimal conditions will depend on numerous parameters. Electrolytic or cathodic H-loading is performed in a solution, and the key parameters are the voltage and current, the solution composition including the acid concentration and the presence of an inhibiter that hinders the recombination of atomic H into $H_2$, and the temperature at which the loading is performed. Other parameters could also be critical – for instance even using Pt as a counter electrode can lead its dissolution and deposition of Pt on the cathode (Rong Chen et al., 2017), the pH of the solution may need careful monitoring as it can be modified by dissolved species from the atmosphere and might require purging by $N_2$ or Ar for instance (Anantharaj & Noda, 2022).

For gaseous H-loading, the temperature, the nature of the gas, and, importantly, its purity (Fromm & Uchida, 1987) and pressure all matter. Numerous designs of cells enabling gas thermochemical treatments of atom probe specimens have been proposed (Bagot et al., 2006; Dumpala et al., 2014; Haley et al., 2017; Perea et al., 2017; Khanchandani, El-Zoka, et al., 2022). Higher temperatures can be achieved through resistive heating, or recently by using a DC laser, which offer the advantages of temperature control allowing for complex heat treatment schedules, including very fast cooling that can be difficult to achieve through resistive heating. Cryo-cooling may assist in capturing sensitive transient (El-Zoka et al., 2023), but can result in the condensation of contamination on the specimen that can preclude further analysis without additional cleaning as is performed after TEM observation for instance (Herbig &

Kumar, 2021). Ultimately, precise calibration of the temperature through thermocouples is typically necessary, particularly as pyrometers often neglect to account for non-ideal emissivity of the targeted sample, with a surface state that can have changed during exposure to the gas.

Finally, with regards to ion implantation, the dose and dose rates are important, but so is the surface state as it had been reported for instance that C-based species can be carried into the material during ion implantation (Dagan et al., 2015) as well as the overall cleanliness of the vacuum inside the implanter (Wang et al., 2017).

### 3.2.2 Loading efficacy

There is overall a need to check the efficacy of the H-loading process, in order to optimize the conditions. For instance, Takahashi et al. noted that H-loading at a higher temperature can favor uptake (Takahashi et al., 2010, 2012). Depending on the material, H permeation through a surface oxide layers can be very difficult, slow or near impossible (Evers et al., 2013). The formation of atomic H may need to be assisted or catalyzed by other metals. Dr. M. Rohwerder (MPISUSMAT) mentioned the use of a thin film of Pd deposited onto the sample of interest when doing Kelvin probe experiments for instance (Lupu et al., 2004; Kesten et al., 2002; Gemma et al., 2009). Coatings deposited by using the approach of Schwarz and co-workers (Schwarz et al., 2024) can be made to cover the entire specimen or only a part of the surface. This opens an opportunity for depositing Pd on a section of the specimen to favor the formation of atomic H on the surface and possibly facilitate H- or D-loading.

In addition, cathodic H-loading has most often been reported on specimens prepared from electrochemical polishing. A site-specific specimen prepared by FIB lift-out presents additional challenges. The weld is typically made of a Pt-C or W-C composite, with a density that depends on the deposition conditions (Felfer et al., 2012). The volume of material to be analyzed is slow, and with a high surface-to-volume ratio. Both the weld and the material of interest can corrode fast and be completely lost (Khanchandani, Kim, et al., 2022). The thickness of the welds must be adjusted, and, to increase the success rate, the sample should be loaded before the final sharpening is performed (Khanchandani, Kim, et al., 2022), preferably at low temperature as discussed above. The possibility of using redeposition welding followed by a metallic reinforcement should be explored in the future (Douglas et al., 2023; Woods et al., 2023).

The high diffusivity of H, and to a lesser extent D, can lead to a loss by outward diffusion and possible recombination and desorption on the surface (Figure 1B), facilitated by the small size of the needle-shaped APT specimens (Haley et al., 2014; Gemma et al., 2009). This makes the transfer time/temperature/pressure critical parameters that need to be reported to facilitate reproducibility of results. For instance, exposure of samples to ambient atmosphere can lead to possible oxidation or corrosion reactions (with moisture) on the freshly prepared specimen surface. This can lead to the release of atomic hydrogen, from the splitting of water, that can then penetrate into the material (Rodrigues & Kirchheim, 1983; Gråsjö et al., 1995). The opposite process could also take place, with the formation of water on the surface that could pick up H on the surface and favor a depletion of H from within the specimen. This water could then evaporate under ambient conditions for instance. Such a reaction could lead to exchanges between H and D in the near surface region, from an exchange with H from residual moisture, lowering the relative amplitude of the D signal. At low temperatures, the formation of ice could act as a barrier, or simply limit the kinetic of these reactions.

*3.2.3 Cryogenic cooling*

Cryogenically cooling specimens immediately after H-loading can slow the outward diffusion of H or D (Takahashi et al., 2018; Y.-S. Chen et al., 2017; Takahashi et al., 2010) and facilitate detection. This however requires dedicated infrastructure (Perea et al., 2017; Stephenson et al., 2018). Using vacuum and cryo-transfer can avoid frosting of cold specimens transported through ambient atmosphere, and protect the reactive surfaces of freshly prepared specimens from the environment. However, the lower H pressure in the surrounding might facilitate outgassing. In addition, as noted above, there is a regime of H-migration by tunnelling that cannot be avoided even by the use of cryogenic transfer. Another consideration is that at a lower temperature, the solubility of H or D into a given phase will likely decrease, Figure 1C, making it possible for hydrides to form for instance or simply changing the partitioning between different phases compared to what it could be at possible higher operating temperatures for materials in service conditions.

Finally, it should be noted that if H and D are typically fast diffusers in e.g. body-centered cubic α-Fe, and if diffusible hydrogen is to be analyzed, then fast cooling following H-loading should be envisaged. This has so far been achieved primarily using liquid nitrogen, but there may be ways to use other coolants as for cryo-TEM (Dubochet, 2016). Chen and co-workers reported losses in the case of the analysis of some carbides (Chen et al., 2019; Y.-S. Chen et al., 2017).

However, this is not always needed, even in ferritic or martensitic steels, since H- or D- can be trapped at microstructural features, for which higher temperatures can be required to induce de-trapping at a sufficiently high rate. This information is typically accessible via TDS for instance. Please note, even in TDS, the detected amount of H is dependent to the size of specimens (Suzuki & Takai, 2012). Ultimately, there have been several reports of analysis of hydrogenated, deuterated or tritiated samples with no cryogenic holding, preparation or transfer leading to successful detection of D / T at microstructural features (Khanchandani et al., 2023; Jakob et al., 2024; Wang et al., 2022; Sun et al., 2021; Devaraj et al., 2021).

## 3.3 Experimental analysis conditions

An important aspect introduced in Section 2.2 is that residual H can be detected in the form of $H^+$, $H_2^+$ and $H_3^+$ (Tsong et al., 1983; Wada et al., 1983; Krishnaswamy & Müller, 1977), with a distribution that exhibits a strong dependence on the electrostatic field conditions, Figure 1B. How these various ionic species form and possibly then dissociate under the effect of the intense electrostatic field (Tsong et al., 1983; Ai & Tsong, 1984) remains an open question, in part because of the complex physics involved (Xu et al., 2017) that will require targeted studies.

To a first approximation, and according to the post-ionization theory (Kingham, 1982), the charge-state ratio can be used to monitor the electrostatic field conditions across datasets, and instruments (Shariq et al., 2009). It should be noted that the actual values of the field derived from the theory may be inaccurate. Examples have been reported where the theory does not readily apply in semiconductors or oxides (Schreiber et al., 2014; Singh et al., 2024; Cuduvally et al., 2022), which may be due to a more complex field evaporation behavior or field-induced dissociations. The charge state of atomic ions can often be used (Kellogg, 1982), but the relative abundance of molecular ions also shows similar trends (Müller et al., 2011), even if their precise formation mechanisms can remain elusive. The ratio of a combination of atomic and molecular ions was used in the case of the analysis of bulk hydrides (Y.H. Chang et al., 2019).

In deuterated samples, overall, the consensus appears to be that when HV pulsing can be used, it should be. HV pulsing leads to higher electrostatic fields and hence reduces the relative fraction of $H_2^+$ and $H_3^+$, making it easier to discriminate D-containing peaks. At higher fields, the relative fraction of multiple events increases, which can lead to more severe ion pile up at the detector and losses of H or D, as discussed in (Y.H. Chang et al., 2019). Laser pulsing facilitates field evaporation at relatively lower electrostatic fields, which promotes the detection of relatively higher levels of residual H and of multiple H-containing species, including possible combinations of D and H that can make peak identification highly complex Figure 2C (Jones et al., 2022; Mouton et al., 2018). A relatively lower electrostatic field however tends to improve yield (Prosa et al., 2019), and using laser pulsing may be the only way to get any data at all.

It is often the case that experiments are run at a constant detection rate, i.e. a fixed average number of ions detected per pulse. However, as the specimen blunts during the experiments, and the emitting area increases, maintaining the detection rate forces a monotonous decrease in the electrostatic field in HV pulsing mode. In laser pulsing mode, the situation can be made more complex as a relative larger specimen volume results in a lower peak temperature for the thermal pulse, which is compensated by a relative increase in the electrostatic field. There is an important optimization process to pursue here, and careful calibrations can hence be the only way to assess if the amount of detected H or D is statistically significant.

## 3.4 Calibrations

For decades, APT has been claimed to be calibration-free, i.e. since the technique relies on counting the number of individual ions of each species. This is expected to contrast with e.g. electron-probe microanalysis (EPMA) or energy-dispersive X-ray spectroscopy (EDS/EDX) for which standards of known compositions are normally necessary for regular calibration. However, this claim is known to be rather wrong – the report already in the 1980s of species-specific losses from the field evaporation at the DC voltage of the element with the relatively lower evaporation field (Miller, 1981). In the time-of-flight spectrum, ions formed from DC-field evaporation, lost from the analysis, combine with the dark current of the detector to form a level of white background. Upon conversion into mass-to-charge ratios, this can result in a dependency of the measured background and composition on the base temperature and pulsing frequency (Hyde et al., 2011; Hatzoglou et al., 2020; Cappelli & Pérez-Huerta, 2023; El-Zoka et al., 2020).

There are other loss mechanisms that can affect the quantitativity of the measurements. First, two or more ions of similar mass-to-charge emitted by a single pulse and flying along similar trajectories can hit the detector in too close proximity for both to be detected (Rolander & Andrén, 1989). This so-called pile-up effect was studied in detail for B (Meisenkothen et al., 2015) and C (Peng et al., 2018) on modern, commercial instruments. Second, dissociation of molecular or cluster ions during the flight can produce neutral atoms or molecules (Gault et al., 2016), with a strong dependence on the electrostatic field conditions (Zanuttini et al., 2017). Tracks in the correlation histogram for multiple events (Saxey, 2011) normally reveal these dissociation reactions, and allow for identifying the corresponding reactions, e.g. $MO_2^+ \rightarrow M^+ + O_2$. To a first approximation, the relative energetic stability of the end products ($M^+ + O_2$) with respect to the parent molecular or cluster ion ($MO_2^+$) provides a guide for estimating if a dissociation can lead to the formation of a neutral atom or molecule (Gault et al., 2016; Blum et al., 2016; Zanuttini et al., 2017, 2018; Kim et al., 2024). Through one or more of these loss mechanisms, compositional inaccuracies have been reported for oxides, carbides, nitrides as well as for metals (Marquis, 2007; Mancini et al., 2014; Müller et al., 2011; Amouyal & Seidman, 2012; Thuvander et al., 2011; Sha et al., 1992; Marceau et al., 2013), with a dependence on the electrostatic field conditions. Ultimately, the question remains if and how these effects impact the analysis of hydrogen by APT.

For analysis of H by APT, one could envisage the analysis of a "standard". This is common practice across fields and for other techniques, including X-ray energy or wavelength dispersive spectroscopy and SIMS. Some standards were also used to compare across techniques (Exertier et al., 2018; Meisenkothen et al., 2015; DeRocher et al., 2022)

During the discussions, it was proposed to use stable hydrides that can be sourced commercially, such as $TiH_2$. Chang and co-workers analyzed a series of specimens prepared from a freshly fractured surface of a stable $TiD_2$ sample (Y.H. Chang et al., 2019), and reported the formation of $D_2$ through the reaction $TiD_2^+ \rightarrow Ti^+ + D_2$ along with multiple events containing two $D^+$ ions that could lead to substantial pile-up. There is also delayed field evaporation that will lead to losses to the background. These mechanisms are all dependent on the amplitude of the electrostatic field, which was monitored by the ratio between charge states ($Ti^{3+}$ and $Ti^{2+}$) or ($Ti^{2+}$ and $TiD_2^+$)

There is another perspective on the needed calibration, which is that the amount of residual H and the corresponding ionic distribution (i.e. the relative amplitudes of $H_{1-3}^+$) can be expected to be reproducible from one experiment to another on a similar instrument, or set of instruments. This was reported for instance for a range of metallic materials as a function of the charge-state ratio of one of the elements within the material (Khanchandani, Kim, et al., 2022; Khanchandani et al., 2023; Freixes et al., 2022; Breen et al., 2020; Meier et al., 2023), and can offer a way to assess whether the measured content of H (or D) falls within the possible range of detectable H in unloaded specimens, or whether the difference is statistically significant. This was also performed on a local basis within the microstructure, since the charge-state ratio is locally accessible (Chang et al., 2018).

Finally, Meier and co-workers (Meier et al., 2022) introduced an alternative calibration by investigating the variations in the detected amount of $H^+$ and $H_2^+/D^+$ as a function of the time in between two pulses. This is based on the assumption that H-migrates along the specimen shank, and hence that the supply of H to the emitting area at the tip of the specimen will be time dependent. A series of experiments were done at varying pulsing frequencies, and detection rates but with the caveat that then the electrostatic field changed. With a shorter time between pulses, the supply of H and hence the detection of $H^+$ and $H_2^+$ should be limited, but the detection of D should not. This approach was used across several materials and data reported and discussed on multiple occasions.

Although such approaches do not provide an actual calibration, they help provide trends as to the levels of H that can be expected under a certain range of electrostatic field conditions. This offers a comparison point as to how far off this trend a particular measurement can be, supporting interpretation of the statistical significance of a locally measured high H- or D-concentration (Breen et al., 2020; Khanchandani, Kim, et al., 2022; Freixes et al., 2022).

## 4  Recommendation for APT data reporting

Blum and co-workers (Blum et al., 2018) proposed a set of information that should be recorded and reported for each APT dataset used in scientific publications focusing on the analysis of geological materials. The table they proposed to use contains critical information to facilitate the analysis of the data by an external expert reader. Similar information was already proposed by Diercks and co-workers in a previous extended abstract (Diercks et al., 2017), as mentioned by Dr. S. Gerstl (SCOPEM, ETH Zurich) during the workshop.

We propose to use a modified and extended version of the table from Blum and co-workers that can be downloaded as a spreadsheet from this link:

https://docs.google.com/spreadsheets/d/1nt5n6uBHsCnv_30RWw--U9lo2vDZW1iWF50LGd2rd7s/edit?usp=sharing

A copy is available in the supplementary information. In putting together this spreadsheet, we tried to align with current efforts in defining an APT-focused ontology to capture relevant metadata to facilitate storage and reuse of data according to the FAIR principles (Wilkinson et al., 2016). The FAIRMAT project, a part of the German's national research data management initiative, appears today to contain the most complete listing available for APT, see link in SI.

We also generally agreed that it would be best to always include a set of additional information in the supplementary material of articles. First, a mass spectrum, with absolute and not relative counts – i.e. not just a normalized spectrum. This will help assess the statistical significance of peaks overall. Ideally, the mass spectrum should be readable, so maybe splitting it into various parts may help. Ideally, the spectrum should be supplied also as a .csv file supplemented by the range file. Second, whenever possible, electron micrographs of specimens should also be included. Ideally, images should be provided at a range of magnifications to assess its shape further down the shank. Third, in the case of H-loading through ion implantation, simulated profiles from e.g. the stopping range of ions in matter, i.e. SRIM (Ziegler & Biersack, 1985) should be included.

We believe that these recommendations could be followed generally across the field, but we propose that all workshop attendees use this from now on to report data in their own scientific articles, and suggest their adoption during peer-review of scientific articles focused on the topic.

## 5  Perspectives and conclusion

In summary, H-analysis by APT is still far from routine, and depends on the question that is to be addressed. These are many effects that compound to make the analysis extremely complex, with numerous parameters that must be accounted for and appropriately reported to ensure reproducibility of the experiments. Figure 1 and Figure 2 summarize some of the key aspects discussed herein and during the workshop.

Whether one aims to measure H in low concentration in solid solution, or segregated amount of H at microstructural features, or finally the composition of hydrides, they will face very different challenges. The first requires low H background, and adjusting the electrostatic field conditions to avoid the detection of $H_2^+$ so D can be confidently labelled and measured. The second also requires a low background level, but also careful analysis of the local field conditions that can change drastically at individual microstructural features. The third appears less problematic, i.e. at least it faces familiar problems of APT in the estimation of stoichiometry of known compounds, with a strong dependence on the electrostatic field conditions. Calibration with analyses carried out on reference samples might be required, especially for the first and third cases, but also a calibration of the expected H contents and distributions as a range of electrostatic field conditions can inform on the statistical significance of local high H or D concentrations. Such a calibration can be made on an unloaded specimens aiming at determining the conditions, i.e. a specific charge-state ratio, that minimizes the influence of residual hydrogen on the analysis.

It seems that the community is, on average, maybe more careful in drawing conclusions from observations of H by APT than it has sometimes been with other elements that are notoriously difficult to quantify. And this is a good thing. Reports have been often limited to qualitative statements, since quantification is extremely challenging. The development of novel instruments with lower levels of residual hydrogen may hold the key for more accurate measurements in the future. We will not be able to perform quantitative analysis without understanding the details of the origins of the signal. Groundwork is therefore needed to go beyond 'the hero experiment' and really reach routine quantitative analyses.

Finally, it remains unclear what is the needed accuracy of a compositional measurement to make it useful. These local H in conjunction with models that could offer predictive power on hydrogen embrittlement susceptibility for instance.

Finally, it remains unclear what is the needed accuracy of a compositional measurement of H to make it useful. These local H content obtained by APT must be combined with various models to advance the understanding of hydrogen embrittlement. For instance, what are the dominant mechanisms between plasticity-mediated hydrogen embrittlement or decohesion mechanism, depending on the local H concentration. When planning such highly demanding APT experiments, it is also important to consider the relevance of measurements performed on H- or D-loaded samples to the hydrogen embrittlement of engineering parts in service operated in completely different atmosphere, pressure, temperature and possibly stress. Nevertheless, APT remains one of the most powerful techniques for H mapping and understanding the behavior of H in complex microstructures.

## Acknowledgements

This workshop was organized and chaired by Aparna Saksena, Baptiste Gault, Xavier Sauvage and Paul Bagot and sponsored by the International Field Emission Society (IFES, http://www.fieldemission.org/). The organizers are grateful for financial support by Thermofisher Scientific, Cameca Instrument and Ferrovac. Apart from the corresponding authors, and the chairs, the authors were listed alphabetically.
Uwe Tezins, Andreas Sturm and Christian Bross are thanked for technical support with the atom probe facility at the MPI SUSMAT. BG acknowledges financial support from the ERC-CoG-SHINE-771602. The APT group at MPI SUSMAT is grateful to the Max-Planck Gesellschaft and the BMBF for the funding of the Laplace Project, the BMBF for the UGSLIT project. BG and AS are grateful for funding from the DFG for SFB TR103/A4, and the CRC 1625/B4. TMS is grateful for funding from the DFG through the award of the Leibniz Prize 2020. JD is grateful for support via EPSRC grant EP/V007661/1. RG thanks to Japan Society for the Promotion of Science (JSPS) KAKENHI via Grant Number JP24K08249. BMJ is a recipient of the WINNINGNormandy Program supported by the Normandy Region and would like to acknowledge that this project has received funding from the European Union's Horizon 2020 research and innovation programme under the Marie Skłodowska-Curie grant agreement No. 101034329. MK acknowledges financial support from the DFG through DIP Project No. 450800666. PF and BO work has received funding from the European Research Council (ERC) under the European Union's Horizon 2020 Research and Innovation Programme (Grant Agreement No. 805065). LD acknowledges support from STFC grants ST/Y004817/1, ST/T002328/1, and ST/W001128/1.

batteries: challenges and ways forward. *Journal of Materials Chemistry A* **6**, 4883–5230.

KIM, S.-H., BHATT, S., SCHREIBER, D. K., NEUGEBAUER, J., FREYSOLDT, C., GAULT, B. & KATNAGALLU, S. (2024). Understanding atom probe's analytical performance for iron oxides using correlation histograms and ab initio calculations. *New Journal of Physics* **26**, 033021.

KINGHAM, D. R. (1982). The post-ionization of field evaporated ions: A theoretical explanation of multiple charge states. *Surface Science* **116**, 273–301.

KOLLI, R. P. (2017). Controlling residual hydrogen gas in mass spectra during pulsed laser atom probe tomography. *Advanced Structural and Chemical Imaging* **3**, 10.

KREMER, K., SCHWARZ-SELINGER, T. & JACOB, W. (2021). Influence of thin tungsten oxide films on hydrogen isotope uptake and retention in tungsten – Evidence for permeation barrier effect. *Nuclear Materials and Energy* **27**, 100991.

KRISHNASWAMY, S. V. & MÜLLER, E. W. (1977). Metal Hydrides in Pulsed Field Evaporation. *Zeitschrift für Physikalische Chemie* **104**, 121–130.

LARSON, D. J., CEREZO, A., JURASZEK, J., HONO, K. & SCHMITZ, G. (2009). Atom-Probe Tomographic Studies of Thin Films and Multilayers. *Mrs Bulletin* **34**, 732–737.

LARSON, D. J., FOORD, D. T., PETFORD-LONG, A. K., LIEW, H., BLAMIRE, M. G., CEREZO, A. & SMITH, G. D. W. (1999). Field-ion specimen preparation using focused ion-beam milling. *Ultramicroscopy* **79**, 287–293.

LI, K., AARHOLT, T., LIU, J., HULME, H., GARNER, A., PREUSS, M., LOZANO-PEREZ, S. & GROVENOR, C. (2019). 3D-characterization of deuterium distributions in zirconium oxide scale using high-resolution SIMS. *Applied Surface Science* **464**, 311–320.

LIU, J., TAYLOR, S. D., QAFOKU, O., AREY, B. W., COLBY, R., EATON, A., BARTRAND, J., SHUTTHANANDAN, V., MANANDHAR, S. & PEREA, D. E. (2022). Visualizing the Distribution of Water in Nominally Anhydrous Minerals at the Atomic Scale: Insights From Atom Probe Tomography on Fayalite. *Geophysical Research Letters* **49**, e2021GL094914.

LIU, P.-Y., ZHANG, B., NIU, R., LU, S.-L., HUANG, C., WANG, M., TIAN, F., MAO, Y., LI, T., BURR, P. A., LU, H., GUO, A., YEN, H.-W., CAIRNEY, J. M., CHEN, H. & CHEN, Y.-S. (2024). Engineering metal-carbide hydrogen traps in steels. *Nature Communications* **15**, 724.

LUPU, D., RADU BIRIȘ, A., MIȘAN, I., JIANU, A., HOLZHÜTER, G. & BURKEL, E. (2004). Hydrogen uptake by carbon nanofibers catalyzed by palladium. *International Journal of Hydrogen Energy* **29**, 97–102.

LYNCH, S. (2019). Discussion of some recent literature on hydrogen-embrittlement mechanisms: addressing common misunderstandings. *Corrosion Reviews* **37**, 377–395.

MAIER, H., SCHWARZ-SELINGER, T., NEU, R., GARCIA-ROSALES, C., BALDEN, M., CALVO, A., DÜRBECK, T., MANHARD, A., ORDÁS, N. & SILVA, T. F. (2019). Deuterium retention in
27